\documentclass[prb,showpacs,graphicx,floatfix,preprint]{revtex4}%
\usepackage{epsf}
\usepackage{amsmath}
\usepackage{amsfonts}
\usepackage{amssymb}
\usepackage{graphicx}%
\begin{document}
\title{Nonanalytic corrections to the specific heat and susceptibility of a non-Galilean-Invariant 
Two-Dimensional Fermi Liquid}
\author{Andrey V. Chubukov$^{1}$ and Andrew J. Millis$^{2}$}
\affiliation{$^1$Department of Physics, University of 
Wisconsin-Madison, 1150 Univ. Ave., Madison WI 53706-1390\\
$^{2}$Department of Physics, Columbia University,  538 W. 120th St, New York, NY 10027}
\date{\today}

\begin{abstract}
We consider the leading non-analytic 
temperature dependence of the specific heat and 
temperature and momentum dependence of the spin susceptibility 
for two dimensional  fermionic systems with non-circular Fermi surfaces. 
We demonstrate the crucial role played by Fermi surface curvature.
 For a Fermi surface with inflection points,  we demonstrate that
thermal corrections to the uniform susceptibility in $D=2$ change from
$\chi_{s} \propto T$ to $\chi_{s} \propto T^{2/3}$ for generic inflection
points, and to $\chi_{s} \propto T^{1/2}$ for special inflection points along
symmetry directions.  Errors in previous work are corrected. Application of the results
to $Sr_2RuO_4$ is given.

\end{abstract}
\maketitle

\section{Introduction}

L. D. Landau's "Fermi liquid theory" provides a robust and accurate
description of the leading low temperature, long wavelength behavior of a wide
range of systems of interacting fermions in two and three spatial dimensions.
In Landau's original work \cite{Landau57} it was assumed that the temperature
($T$) and momentum ($q$) corrections to the leading Fermi liquid behavior were
analytic functions of $(T/T_{F})^{2}$ and $(q/k_{F})^{2}$, with the Fermi
momentum $k_{F}$ set by the inter-particle spacing and the Fermi temperature
$T_{F}\sim v_{F}k_{F}$ with $v_{F}$ a typical measured electron velocity.
However, subsequent work revealed that in dimensions $d=2$ and $d=3$
the leading temperature and momentum dependences of 
measurable quantities including the  specific heat and spin susceptibility
are in fact non-analytic functions of $T^{2}$ and $q^{2}$.  The nonanalyticities have been studied in detail for Galilean-invariant systems (spherical Fermi surface or circular Fermi line)
(see Ref ~\onlinecite{Chubukov06} for a list of references).  In this paper we extend the
analysis to a more general class of systems, still described by Fermi liquid theory
but with a Fermi surface of arbitrary shape. 

The extension is of interest in order to allow comparisons to systems
such as $Sr_2RuO_4$ \cite{Bergemann03} and quasi-one-dimensional
organic conductors \cite{organic}, in which lattice effects are important. 
The extension also provides
further insight into a fundamental theoretical issue: a crucial finding
of the previous analysis \cite{Chubukov05,Chubukov06}  was that for
two dimensional systems the non-analytical term 
in the specific heat coefficient $\delta C(T)/T \propto T$
arise solely from "backscattering" processes
 at any strength of fermion-fermion interaction.
For the spin susceptibility, the situation is more complex: in addition
to the backscattering contributions to 
$\delta \chi_s (T) \propto T$ and $\delta \chi_s (q) \propto q$  there are
 contributions
of third and higher order in the interaction which
involve non-backscattering processes \cite{ch_masl_latest}. If
the interaction is not too strong the backscattering terms dominate.

Unlike scattering at a general angle, 
the kinematics of backscattering 
is effectively one-dimensional, and depends sensitively
on the shape of the Fermi surface. 
We shall show that the crucial parameter is the Fermi surface
curvature, and that for two dimensional systems in which the Fermi surface 
possesses an inflection point,
the power laws change from $\left|(T,q)\right|$ to   $\left|(T,q)^{1/2}\right|$ or $\left|(T,q)^{2/3}\right|$ 
according to whether the inflection point is or is not along a reflection symmetry axis of the material.
Similar effects occur in three dimensional systems, but 
the effects will be weaker as the singularities there are only logarithmic.

The
importance of the Fermi surface geometry was previously noted by Fratini and
Guinea, who showed that the presence of inflection points changes the power
law for the spin susceptibility $\chi_s (T)$ \cite{Fratini02}.  
However, 
we believe that their calculation 
treated the kinematics of the backscattering incorrectly; as
a result, they found that, in $2D$, the presence
of the inflection point changes the temperature dependence 
only by a logarithm, from $\chi_s (T) \propto T$ to $\chi_s (T) \propto T \log T$,
instead of the $T^{\left(\frac{1}{2},\frac{1}{3}\right)}$ found here.
  
The  paper is organized as follows. Section II introduces the
model and defines notation.  In Section III, we demonstrate  the physical origin of the results,
via a calculation of the long range dynamical correlations of a Fermi liquid.
Section IV presents results for the specific
heat of a multiband system.
In Section V we calculate the nonanalyticities
in the momentum and temperature dependence of $\chi$.
Section VI shows how new power laws emerge for Fermi surfaces with
inflection points, and why one-dimensionality affects the powers.
Section VII presents estimates of the size of the nonanalytic terms in $Sr_2RuO_4$. 
Section VIII is a conclusion.
 
\section{Model}

\begin{figure}[ptb]
\centerline{\epsfxsize=4in \epsfbox{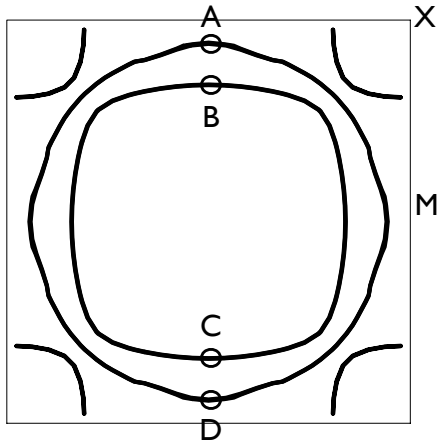}}
\caption{Fermi surface of $Sr_2RuO_4$ computed from
tight binding parameters deduced from quantum oscillation measurements
\cite{Bergemann03}, 
with a few "parallel tangents" points indicated by letters A,B,C,D and the Brillouin
zone points $M=(\pi,0)$ and $X=(\pi,\pi)$ also noted.}%
\label{fsfig}
\end{figure}

We study fermions moving in two dimensions in a  periodic potential. 
As discussed at length in Ref. [\onlinecite{Chubukov06}], because we are
concerned with the low $T$ properties of a Fermi liquid we may adopt a quasiparticle
picture. Lifetime effects
are not important, and the quasiparticle weight ($z$) factors may be absorbed in
interaction constants.  We may therefore consider several bands, labeled by band index $a$,
of quasiparticles moving with (renormalized) dispersion $\varepsilon_{p}^a$.
The Fermi surfaces are defined by the condition
$\varepsilon_{\mathbf{k}}^a=\mu$; an example is shown in Fig \ref{fsfig}.
We parameterize the position at the Fermi surface by a coordinate $s^a$. For vectors $\mathbf{k}$ near a particular
Fermi surface point $\mathbf{k}_{F}$ we will write%
\begin{equation}
\varepsilon_{\mathbf{k}}^a-\mu=v_{F}^a(s^a)\left(  k_{\Vert}+\frac{k_{\bot}^{2}}%
{2k_{0}(s^a)}\right)  \label{ep}%
\end{equation}
with $v_{F}(s^a)=\left\vert \partial\varepsilon_{\mathbf{k}}^a/\partial
\mathbf{k}\right\vert $ the Fermi velocity at the point $(s^a)$ 
and the components of $\mathbf{k}$
parallel and perpendicular to the Fermi velocity (Fermi surface normal) 
given by

\begin{align}
k_{\Vert} &  =\left(  \mathbf{k}-\mathbf{k}_{F}\right)  \cdot\widehat
{\mathbf{v}}_{F}(s^a)\label{ppar}\\
k_{\bot} &  =\left(  \mathbf{k}-\mathbf{k}_{F}\right)  \times\widehat
{\mathbf{v}}_{F}(s^a)\label{pperp}%
\end{align}
$k_0^{-1}(s^a)$ is the curvature of the Fermi surface at the point $s^a$. 
For a circular Fermi surface, $k_0=k_F$ independent of $s$; but in
general  $k_0\neq k_F$ and both depend on $s$.

It is sometimes convenient to use the variables $\varepsilon_k$
and $\theta_k$, where $\theta_k$ is the angle determining the direction of the  Fermi velocity
${\bf v}_F(k)=\partial \varepsilon_k/\partial {\bf k}$ relative to some fixed axis. Eq
\ref{ep} shows that the Jacobean
of the transformation is
\begin{equation}
d^2k=\frac{k_0(k)}{v_F(k)}d\varepsilon_k d\theta_k.
\label{jacobean}
\end{equation}

In a non-Galilean-invariant system the  Fermi surface
may contain   {\em inflection points} at which the curvature vanishes,
i.e. $k_0\rightarrow \infty$. If the inflection point does not lie on an  axis
of reflection symmetry
of the Brillouin zone,  the dispersion (measured in terms of the difference
of the momentum from an inflection point) is 
\begin{equation}
\varepsilon_{\mathbf{k}}-\mu=v_{F}\left(  k_{\Vert}+\frac{k_{\bot}^{3}}%
{k^2_{1}}\right)  \label{ep_1}%
\end{equation}
However, if the inflection point lies on a symmetry axis, then only even 
powers in $k_\bot$ may occur and 
\begin{equation}
\varepsilon_{\mathbf{k}}-\mu=v_{F}\left(  k_{\Vert}+\frac{k_{\bot}^{4}}%
{k^3_{2}}\right)  \label{ep_11}%
\end{equation}
Here $k_1$ and $k_2$ are coefficients  expected in general to be $\sim k_F$.


The fermions interact. We assume that the $T\rightarrow 0$, long wavelength properties
are described by the Fermi liquid   theory,
 so that at low energies the interactions may
be parameterized by the fully reducible Fermi surface to Fermi surface
scattering amplitude 
$\Gamma_{\alpha,\beta,\gamma\delta}(k,p;k,p)$.
For particles near the Fermi surface, 
$|{\bf k}|, |{\bf p}| \approx k_F$, and $\Gamma$ depends on the angle 
$\theta$ between ${\bf k}$ and ${\bf p}$ and on the band indices
$a$ (for $k$) and $b$ (for $p$). Backscattering corresponds to $\theta = \pi$.
 It is often useful to decompose
 $\Gamma_{\alpha,\beta;\gamma,\delta}(\pi)$ into charge and spin components
\begin{equation}
\Gamma^{ab}_{\alpha,\beta;\gamma,\delta}(\pi) = 
\Gamma^{ab,c} \delta_{\alpha\gamma}\delta_{\beta \delta} 
+ \Gamma^{ab,s} {\bf \sigma}_{\alpha\gamma}{\bf \sigma}_{\beta \delta}
\label{m13_1}
\end{equation}

It is 
 also instructive to make contact with second order perturbation
theory for a model in which  the particles are subject to a spin-independent interaction
\begin{equation}
H_{int}=\sum_{\mathbf{q}}U(\mathbf{q})\rho_{\mathbf{q}}\rho_{-\mathbf{q}%
}\label{Hint}%
\end{equation}
with charge density operator
\begin{equation}
\rho_{\mathbf{q}}=\sum_{\mathbf{p}\alpha}c_{\mathbf{p+q},\alpha}^{\dag
}c_{p,\alpha}\label{rho}%
\end{equation}
The leading perturbative result is then 
\begin{equation}
\Gamma^c =  U(0) -\frac{U(2k_{F})}{2}, ~~~\Gamma^s = - \frac{U(2k_{F})}{2} 
\label{m13_2}
\end{equation} 

If the interaction is local and only one band is relevant, i.e., $U(q) = U$, $\Gamma^c = -\Gamma^s = U/2$,
and
\begin{equation}
\Gamma_{\alpha,\beta;\gamma,\delta}(\pi) = 
U  \left(\delta_{\alpha\gamma}\delta_{\beta \delta} -
\delta_{\alpha\delta}\delta_{\beta \gamma}\right)
\label{new_1}
\end{equation}

\section{Physical Origin of Nonanalyticities; role of curvature}

Previous studies of  the isotropic case demonstrated~\cite{Chubukov05,Chubukov06} that
the nonanalyticities in the specific heat and spin susceptibility arise from the long spatial range
dynamical correlations characteristic of  Fermi liquids.
These are of two types.
One involves slow ($|\Omega|<v_Fq$) long wavelength fluctuations  and is expressed
mathematically in terms of the long wavelength limit of the polarizibility
$\delta\Pi_{LW}\equiv lim_{q\rightarrow 0}\Pi(q,\Omega)-\Pi(q,0)\sim \left|\Omega\right|/q$. 
 The $1/q$ behavior of polarizibility gives rise to 
a long-range correlation between fermions which decays as $|\Omega|/r$ at
 distances $1 \ll rk_F < E_F/|\omega|$. 
The other involves processes with momentum transfer 
$q \approx  2p_F$, and is  
expressed mathematically in terms of the polarizibility 
$\delta \Pi \equiv \Pi(q,\Omega)-\Pi(2k_F,\Omega)\sim \Omega/
\sqrt{2k_F-q}$ (for $|\Omega| <2k_F-q$.  The $1/\sqrt{2k_F-q}$ behavior of $\delta \Pi$ 
gives rise to an oscillation with a slowly decaying
envelope, $\cos(2k_{F} r-\pi/4) |\Omega|/\sqrt{r}$ , 
again  at distances $1 \ll rp_F < E_F/|\omega|$ and again leading to singular behavior.
We now compute these processes
in the multiband model defined above,
and then show how they affect thermodynamic variables. 

Consider the long
wavelength process first. The non-analyticity comes from a particle-hole pair excitation
in which both particle and hole are in the same band, and have momenta
in  the vicinity of Fermi surface points 
$s_a^*$ satisfying $\vec{v}_F(s_a^*)\cdot {\vec q}=0$.
 Choosing one of these points as the origin of
coordinates we have
\begin{eqnarray}
\delta \Pi^{aa}_{s_a^*}&\equiv&\Pi_{s_a^*}(q,\Omega)-\Pi_{s_a^*}(q,0)=T\sum_n\int \frac{dk_\Vert dk_\bot}{(2\pi)^2}\frac{1}{i\omega_n-v_F(s_a^*)\left(k_\Vert+\frac{(k_\bot -q/2)^2}{2k_0(s_a^*)}\right)}
\nonumber\\
&\times&\left(\frac{1}{i\omega_n+i\Omega_n-v_F(s_a^*)\left(k_\Vert+\frac{(k_\bot+q/2)^2}{2k_0(s_a^*)}\right)}-\frac{1}{i\omega_n-v_F(s_a^*)\left(k_\Vert+\frac{(k_\bot + q/2)^2}{2k_0(s_a^*)}\right)}\right)
\label{Pi0}
\end{eqnarray}
Performing the integral over $k_\Vert$ and the Matsubara sum as usual yields
\begin{equation}
\delta \Pi^{aa}_{s_a^*}=\int \frac{dk_\bot}{2\pi}\frac{1}{2\pi v_F(s_a^*)}\frac{i\Omega}{i\Omega-\frac{v_F(s_a^*) k_\bot q}{k_0(s_a^*)}}
\end{equation}
We see that for the part of $\delta \Pi$ which is even in $\Omega$
the integral  is indeed dominated
by
 $k_\bot \sim \Omega k_0(s_a^*)/v_F(s_a^*)$,
 so the approximation of expanding near
this point is justified, and we obtain the nonanalytic long wavelength contribution
as a sum over all Fermi points $s_a^*$ satisfying $\vec{v}_F(s_a^*)\cdot\vec{q}=0$
with coefficients determined by the local Fermi velocity and local curvature:
\begin{equation}
\delta \Pi^{aa}_{LW}(q,\Omega)=\frac{\left|\Omega\right|}{q}\left(\sum_{s_a^*}\frac{k_0(s_a^*)}{4\pi v_F^2(s_a^*)}\right)
\label{pilong}
\end{equation}
For a circular Fermi surface, two points satisfy 
$\vec{v}_F (s_a^*)\cdot\vec{q}=0$, $k_0=k_F$
and Eq. (\ref{pilong}) \ reduces to
the familiar result $k_F |\Omega|/(2 \pi v^2_F q)$. 
If the curvature vanishes, then use of Eqs (\ref{ep_1}) or (\ref{ep_11})
 in Eq (\ref{Pi0}) 
yields a $\delta \Pi_{LW}\sim\left( \Omega/q\right)^{1/2,1/3}$ respectively.

We next consider the "$2k_F$" process. Here the situation is a little different.
The singularities in general come from processes connecting two Fermi points with
parallel tangents (the importance of parallel tangents points has been noted in other contexts
\cite{Altshuler95})--for example the points 
$A,B$ or $A,C$ shown in Fig \ref{fsfig}.
For a given vector ${\bf q}$ we denote as ${\bf Q}$ 
the closest vector  which is
parallel to ${\bf q}$ and
connects two "parallel tangents" points. Symmetry
ensures that the leading dependence of $\delta \Pi^{ab}_Q=\Pi({\bf q})-\Pi({\bf Q})$ involves only
the $q_{\Vert} =({\bf q}-{\bf Q})\cdot {\bf Q}/|{\bf Q}|$. 
Labeling the initial and final
of the  two Fermi points connected
by $Q$ as $s_{1,2}$ and noting that for systems with inversion symmetry the points come in
pairs symmetric under interchange to the band indices we have
\begin{eqnarray}
&& \delta \Pi^{ab}_Q \equiv \Pi({\bf q},\Omega)-\Pi({\bf Q},0)=T\sum_n\int \frac{dk_\Vert dk_\bot}{(2\pi)^2}
\frac{1}{i\omega_n-v_F(s_1)\left(k_\Vert+\frac{k_\bot^2}{2k_0(s_1)}\right)} \times
\nonumber\\
&&\left(\frac{1}{i\omega_n+i\Omega_n-v_F(s_2)\left(k_\Vert+q_\Vert+\frac{k_\bot^2}{2k_0(s_2)}\right)}-\frac{1}{i\omega_n-v_F(s_2)\left(k_\Vert+\frac{k_\bot^2}{2k_0(s_2)}\right)}\right)+(1 \leftrightarrow 2)
\label{PiQ}
\end{eqnarray}
Here the {\em sign} of $v_F$ and $k_0$ becomes important. For two points
"on the same side" of the Fermi surface (e.g. points $A$ and $B$ in Fig \ref{fsfig})
the two velocities and the two curvatures  have the same sign (a change of $k_\Vert$ either increases or decreases
both energies),
the integrations proceed as in the analysis of Eq (\ref{Pi0}),
 and the different position of the $q$ 
means that to the order of interest there is no singular non-analytical term.
On the other hand, if the two velocities have opposite sign (e.g. points $A$, $C$ in 
Fig \ref{fsfig}) then after integrating over $k_\Vert$ and $k_\bot$ we obtain (for $\Omega>0$)
\begin{equation}
\delta \Pi_Q=\frac{\sqrt{k_{avg}}}{\left|4 v_{F1}v_{F2}\right|}T\sum_{\omega_n>0 \hspace{.02in} or \hspace{0.02in} \omega_n<-\Omega}
\frac{sgn(\omega)}{\sqrt{\frac{2i\omega}{v_{avg}}+\frac{i\Omega}{v_{F2}}-q_\Vert}} +\left(1\leftrightarrow2\right)
\end{equation}
Here 
$v_{F1} = v_F (s_1), ~v_{F2} = v_F (s_2)$, and 
\begin{eqnarray}
\frac{1}{k_{avg}}&=&\frac{1}{2}\left(\frac{1}{k_0(s_1)}+\frac{1}{k_0(s_2)}\right)\\
\frac{1}{v_{avg}}&=&\frac{1}{2}\left(\frac{1}{\left|v_{F1}\right|}+\frac{1}{\left|v_{F2}\right|}\right)
\label{vavg}
\end{eqnarray}
Completing the evaluation yields
\begin{equation}
\delta \Pi_Q= \frac{\sqrt{k_{avg}}}{4\pi \left|( v_{F1}\right| +
\left| v_{F2}\right|)}
~\left(\sqrt{q_\Vert + \frac{i\Omega}{|v_{F1}|}} + 
 \sqrt{q_\Vert - \frac{i\Omega}{|v_{F1}|}}+1\leftrightarrow2\right)
\label{new_2}
\end{equation}
The singular $|\Omega|/\sqrt{q}$ 
behavior of the dynamic $\delta \Pi_Q$ only holds
 when $q_{\Vert} <0$ and is obtained by expanding (\ref{new_2}) 
 in $\Omega/q_{\Vert}$. On the contrary, the singular behavior of the static 
$\Pi_Q \propto \sqrt{q_{\Vert}}$ holds  at $q_{\Vert} >0. $
The dynamic part of $\Delta \Pi_Q$ behaves as $\Omega^2/(q_{\Vert})^{3/2}$
 at $q_{\Vert} >0$.

\begin{figure}[ptb]
\centerline{\epsfxsize=2in \epsfbox{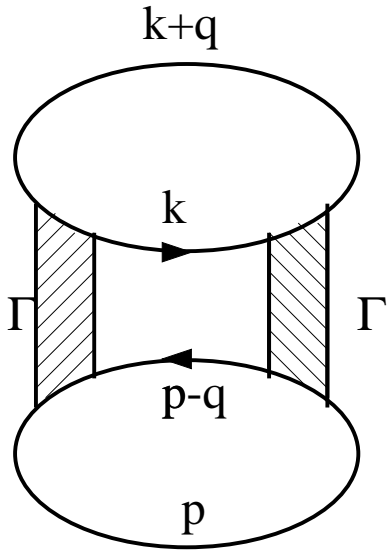}}
\caption{
One of the two  second-order diagrams for the free energy,
which give the non-analytic contribution 
to the specific heat (from Ref.\protect\onlinecite{Chubukov06}). 
 Here $\Gamma$ 
is the (unsymmetrized) fully renormalized Fermi surface to Fermi surface scattering amplitude.} \label{Omega}%
\end{figure}

\section{Specific Heat}
 
This section treats the non-analyticity in the specific heat.  The 
Galilean-invariant case studied previously  is simple enough that the 
corrections can be evaluated, with no
approximations beyond the usual ones of Fermi Liquid theory \cite{Chubukov06}.
The evaluation confirms that in two dimensions the non-analytical contributions to the specific heat 
involve only the backscattering amplitude 
 $\Gamma(\pi)$.  In work prior to that reported in Ref [\onlinecite{Chubukov06}]
 this conclusion was reached
by approximate calculations in which  it was assumed that the non-analytical contributions
were governed by the backscattering amplitude only, and then the assumption was shown
{\em a fortiori} to be consistent.  In the non-Galilean-invariant case of interest here, 
a complete analysis along the lines of Ref [\onlinecite{Chubukov06}]
is not possible. We will follow earlier work and assume that the effects
arise only from backscattering processes, and then show that the assumption is self consistent. We specialize
for ease of writing to a momentum-independent vertex 
$\Gamma$ (see (\ref{new_1})), but  keep band indices. 
For a two-dimensional system
the diagram which gives  the nonanalytic term in the thermodynamic potential $\Xi$  
is \cite{Chubukov06} shown in Fig \ref{Omega}. We may write the resulting diagram schematically as
\begin{equation}
\Xi=
-\frac{1}{2} 
\sum_{abcd}\int (dqd\Omega)\Gamma_{abcd}^2\Pi_{ab}(q,\Omega)\Pi_{cd}(q,\Omega)
\label{xi}
\end{equation}
Now, singularities leading to non-analytical terms may arise for $q\rightarrow0$, in which case
we must consider only the intraband contribution to $\Pi$, and $q\rightarrow Q$, where $Q$
is one of the "parallel tangents" vectors mentioned in the previous section, in which case the
band indices of polarizabilities may be different. 
Let us consider first the small $q$ singularities.   We have
\begin{equation}
\Xi_{LW}=
-\frac{1}{2} 
\sum_{ab}\int (dqd\Omega)\Gamma_{aabb}^2\Pi_{aa}(q,\Omega)\Pi_{bb}(q,\Omega)
\label{xi1}
\end{equation}
Substituting from Eq (\ref{pilong}) gives
\begin{equation}
\Xi_{LW}=
-\frac{1}{2} 
T\sum_\Omega\int \frac{qdqd\theta}{(2\pi)^2}\sum_{ab}\Gamma_{aabb}^2\frac{\Omega^2}{q^2}\frac{k_0^a(\theta)k_0^b(\theta)}{(4\pi)^2 v_{F,a}^2v_{F,b}^2}
\label{chilong}
\end{equation}
The integral over $q$ is logarithmic and is cut by $\Omega$; the analytical continuation
and integral of frequencies may then be performed and we obtain
\begin{equation}
 \left.\frac{\delta C}{T}\right|_{LW}=- \frac{3\zeta(3)}{\pi^3}
\sum_{ab}\int\frac{d\theta}{2\pi}
\frac{\Gamma_{aabb}^2k_0^a(\theta)k_0^b(\theta)}{v_{F,a}^2v_{F,b}^2}
\end{equation}

We now turn to the parallel tangents part of the calculation, finding
\begin{equation}
\Xi_Q=
-\frac{1}{2}
\sum_{ab}\int (dqd\Omega)\Gamma_{abba}^2\Pi_{ab}(Q+q,\Omega)\Pi_{ba}(Q+q,\Omega)
\label{xiq}
\end{equation}
Note that the symmetry of the vertex means that $\Gamma_{abba}^2=\Gamma_{aabb}^2$.
Again substituting 
$\Delta \Pi_Q$ instead of $\Pi_{ab}(Q+q,\Omega)$
and evaluating the integrals explicitly we find
~\cite{comm_2} 
(note that to obtain the non-analytical behavior
it is sufficient to expand $\delta \chi_Q$ for $|\Omega| <<q$)
\begin{equation}
\Xi_Q=
-\frac{1}{2} 
\sum_{ab}\Gamma_{abba}^2T\sum_\Omega\int \frac{dq_\Vert dq_\bot}{(2\pi)^2}\frac{\Omega^2}{(4\pi)^2 v_{Fa}^2v_{Fb}^2}\frac{k_{avg}}{|q_\Vert|}
\label{xiq2}
\end{equation}
Now noting that $k_{avg}=k_0^ak_0^b/(k_0^a+k_0^b)$ and that   $dq_\bot=(k_0^a+k_0^b)d\theta$
we see that $\Xi_Q$ and $\Xi_{LW}$ give identical contributions despite the apparently different 
kinematics.

Adding the two contributions, we obtain 
\begin{equation}
 \frac{\delta C}{T} =- \frac{3\zeta(3)}{2 \pi^3}
\sum_{ab}\int\frac{d\theta}{2\pi}
\frac{\Gamma_{aabb}^2k_0^a(\theta)k_0^b(\theta)}{v_{F,a}^2v_{F,b}^2}
\label{apr4_1}
\end{equation} 
Note that the integrals over the Fermi surface contain $k^2_0 = k^2_0 
(\theta_k)$ rather than the product of two $k_0$ factors at different points along the Fermi surface. 
This is a direct consequence of the fact that only backscattering contributes to  (\ref{eq32}) and (\ref{eq33}).
For a one band model with an isotropic Fermi surface, Eq. (\ref{apr4_1}) 
reduces to the result in \cite{Chubukov05}.   For a generic interaction, 
the calculation goes through as before with
$\Gamma^2_{aabb}$  replaced the components of the
fully renormalized, symmetrized  
Fermi surface to Fermi surface backscattering amplitude so that
 \begin{equation}
 \frac{\delta C}{T} =- \frac{3\zeta(3)}{2 \pi^3}
\sum_{ab}\int\frac{d\theta}{2\pi}
\frac{\left( \Gamma^{ab,c}(\pi)^2  + 3\Gamma^{ab,s}(\pi)\right)^2k_0^a(\theta)k_0^b(\theta)}{v_{F,a}^2v_{F,b}^2}
\label{apr4_1_1}
\end{equation}

%
\section{Susceptibility}

\begin{figure}[ptb]
\begin{center}
\epsfxsize=0.8\columnwidth \epsffile{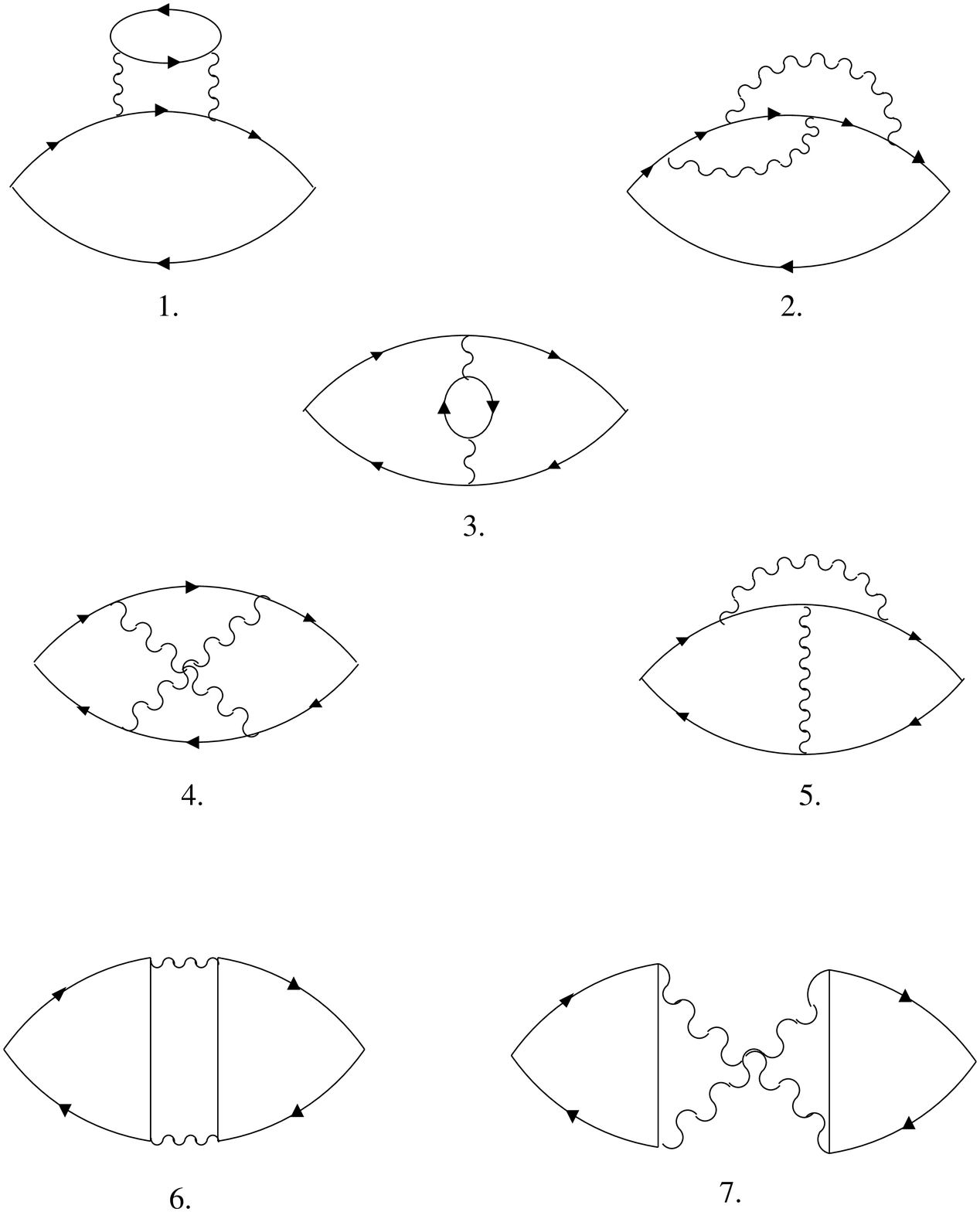}
\end{center}
\caption{Relevant second-order diagrams for the spin and charge susceptibilities (ferm Ref.\protect\cite{Chubukov05}).
The last two diagrams are non-zero only for the charge susceptibility.}%
\label{diagramsforchi}%
\end{figure}

\subsection{Overview}

This section presents calculations of the non-analytical momentum and temperature dependence of
the spin susceptibility of a  two dimensional non-Galilean-invariant Fermi liquid system. 
with a Fermi surface without inflection points. In contrast to the specific heat, there are two
classes of contributions to the nonalnayticities in the susceptibility \cite{ch_masl_latest}. 
One is of second order in the fullly 
renormalized interaction amplitudes, involves
 only the backscattering, and is treated here. 
The other, which we do not study here, is of third and higher orders, and 
involve averages of the interaction
function over a wide range of angles
 (analogously to similar contributions to the
specific heat of a {\em three} dimensional Fermi liquid \cite{Chubukov06}).
 The former process is dominant at weak coupling, and
involves an integral of the square of the curvature over the Fermi arc. The latter process has a less singular dependence on the curvature. 

Even to the order at which we work,
many diagrams contribute (see Fig \ref{diagramsforchi}); we evaluate one in detail to illustrate
the basic ideas behind the calculation and then simply present
the result for the sum of all diagrams. 
Consider for definiteness the "vertex correction" diagram 
-- diagram 3 in Figure  \ref{diagramsforchi}. 
The analytical expression corresponding to this diagram is (we use a condensed notation
in which (dk) stands for an  integral over momentum, 
normalized by $(2\pi)^2$, and a sum over the corresponding Matsubara frequency)
\begin{equation}
\delta\chi(q,0)=-4\sum_{ab} \int(dk)(dl) G^a(k+q)G^a(k)\Lambda^{ab}_k(l) G^a(k+l+q)G^a(k+l)
\label{chibasic}
\end{equation}
and $\Lambda$ is the product of the interaction vertices and internal polarization bubble:
\begin{equation}
\Lambda^{ab}_k(q)=\int(dp)\left(\Gamma^{aabb} (\theta) \right)^2  G^b(p+q/2)G^b(p-q/2)]
\label{lambdadef}
\end{equation}
where $\theta$ is the angle between ${\bf k}$ and ${\bf p}$,
and we have used the fact that 
${\bf k}$ and ${\bf p}$ are near the Fermi surface. 
A complete calculation in the 
Galilean-invariant case shows that,
just as for the specific heat, 
the non-analytical momentum dependence 
of the susceptibility 
arises from
the regions of small $l$ and $l\sim 2k_F$. We consider these in turn. 

\subsection{Small l}
Here we choose a particular point $s_k$ on the Fermi surface and
integrate over $\varepsilon_k$ and the corresponding  frequency. 
 We parametrize the position on the Fermi surface
by the angle $\phi$ between ${\mathbf v}_k$ and ${\bf q}$ and use
Eq \ref{jacobean}. We adopt
coordinates $l_\Vert$ and $l_\bot$ denoting directions parallel and perpendicular to 
${\bf v}(s_k)$ and obtain
 \begin{eqnarray}
\delta \chi_{LW}(q)&=& - \frac{4}{ 2\pi }\sum_{ab} \int \frac{d\theta_k k^a_0 (\theta_k)}{2\pi (v^a_k)^2}
\int \frac{dl_\Vert dl_\bot}{(2\pi)^2}T\sum_\Omega
\frac{i\Omega}{\left({\bf v}^a_k\cdot {\bf q}\right)^2} \Lambda^{ab}_k(l_\Vert,l_\bot,\Omega)
\nonumber \\
&&\left(\frac{1}{\frac{i\Omega}{v^a_k}-{\hat {\bf v}}_k\cdot {\bf q}-l_\Vert}+\frac{1}{\frac{i\Omega}{v^a_k}
+{\hat {\bf v}}_k\cdot {\bf q}-l_\Vert}-\frac{2}{\frac{i\Omega}{v^a_k}-{\hat{\bf v}}_k\cdot {\bf q}-l_\Vert}\right)
\label{chismallq1}
\end{eqnarray}
We may similarly evaluate $\Lambda$, proceeding from Eq (\ref{lambdadef}).
Choosing as origin the point ${\bf p}=-{\bf k}$, defining coordinates 
 $p_\Vert$ and $p_\bot$ antiparallel
and perpendicular to ${\bf v}_k$, integrating over $p_\Vert$ and the corresponding loop frequency gives
\begin{equation}
\Lambda^{ab}_k(l_\Vert,l_\bot,\Omega)=\frac{\Gamma^{aabb}(\pi)^2}{2\pi (v^b_k)^2}\int\frac{dp_\bot}{2\pi}\frac{i\Omega}{\frac{i\Omega}{v^b_k}  +l_\Vert-\frac{p_\bot l_\bot}{k^b_0}}
\label{Lambda2}
\end{equation}

Viewed as a function of $l_\Vert$ the second line  in Eq (\ref{chismallq1})
decays rapidly ($\sim l_\Vert^3$) at large $l_\Vert$
and has poles only in the half plane $sgn Im l_\Vert=sgn\Omega$.  Evaluation of the $l_\Vert$
integral by contour methods, closing the contour in the half plane $sgn l_\Vert=-sgn\Omega$ shows that
nonanalyticities can only arise from singularities of $\Lambda$. Reference to Eq (\ref{Lambda2}) 
 shows that these can only arise from momenta ${\bf p}$ satisfying ${\bf v}_p\cdot{\bf v}_k<0$. 
In the Galilean-invariant case the integral could be evaluated exactly; the resulting non-analytical
terms were found to be determined by a very small  region around ${\bf p}=-{\bf k}$, 
 i.e., around $\theta = \pi$ in (\ref{lambdadef}). 
Here we assume that this is the case, and show {\it a fortiori} that the assumption is consistent.

Performing the integral over $l_\Vert$ yields
\begin{eqnarray}
\delta \chi_{LW} (q)&=&-4i \sum_{ab}\frac{(\Gamma^{aabb} (\pi))^2}{(2\pi)^2}
 \int\frac{d\theta_{k} k^a_0 (k)}{2\pi (v^a_kv^b_k)^2} T\sum_\Omega 
\int\frac{dl_\bot dp_\bot}{(2\pi)^2}\left(\frac{\Omega}{{\bf v}^a_{k}\cdot {\bf q}}\right)^2 \text{sgn} \Omega
\nonumber \\
&&\left(\frac{1}{\frac{2i\Omega}{v_{avg}}-{\hat {\bf v}}_{k}\cdot {\bf q}-\frac{l_\bot p_\bot}{k^b_0}}
+\frac{1}{\frac{2i\Omega}{v_{avg}}+{\hat {\bf v}}_{k}\cdot {\bf q}-\frac{l_\bot p_\bot}{k^b_0}}
-\frac{2}{\frac{2i\Omega}{v_{avg}}-\frac{l_\bot p_\bot}{k^b_0}}
\right)
\label{chitriad3}
\end{eqnarray}
with $v_{avg}$ defined in Eq \ref{vavg}.
Performing the sum  over frequency and rescaling each of  $l_\bot ,p_\bot$ by $\sqrt{k^b_0}$ yields
\begin{eqnarray}
\delta \chi_{LW}(q)&=&-\sum_{ab} \frac{\left(\Gamma^{aabb} (\pi)\right)^2}{8\pi^3}
\int \frac{d\theta_k k^a_0k^b_0}{2\pi (v^a_k v^b_k)^2}\int_{-\Lambda}^\Lambda 
\frac{dl_\bot dp_\bot}{(2\pi)^2}~\frac{v_{avg}^3}{(v^a)^2 }
\nonumber \\
&&\frac{\left(\Upsilon(l_\bot,p_\bot;{\hat {\bf v}}\cdot {\bf q})
+\Upsilon(l_\bot,p_\bot;-{\hat {\bf v}}\cdot {\bf q})-2\Upsilon(l_\bot,p_\bot;0)\right)}{({\hat{\mathbf v}}_k\cdot {\bf q})^2} 
+...
\label{chismallq2}
\end{eqnarray}
with
\begin{equation}
\Upsilon(x,y;z)=(xy-z)^2 \log |xy-z|
\label{12_1}
\end{equation}

Eq (\ref{chismallq2}) is based on an expansion for small $l_\bot ,p_\bot$. The integrals over these quantities
are cut off by other physics above a cutoff scale $\Lambda$ which we have written as a hard cutoff. The ellipsis
denotes other terms arising from physics at and beyond the cutoff scale, which lead to additional,
regular contributions to $\delta \chi$ involving positive, even powers of $q$.
Evaluation of Eq \ref{chismallq2} yields (details are given in  Appendix~\ref{app:details})
\begin{equation}
\delta \chi_{LW}(q)=-\sum_{ab} \frac{\left(\Gamma^{aabb} (\pi)\right)^2\left|{\bf q}\right|}{
 48\pi^3}
\int \frac{d\theta_k k^a_0 k^b_0}{2\pi (v^a_k)^2(v^b_k)^2}\frac{v_{avg}^3}{(v^a)^2 }
\left|{\hat {\mathbf v}}_k\cdot {\hat q}\right|+...
\label{chi3}
\end{equation}
where $v_k =v_F (\theta_k)$, $k_0 = k_0 (\theta_k)$, 
 ${\hat {\mathbf v}}_k\cdot {\hat q} = \cos (\theta_k - \theta_q)$, and 
$\theta_q$ is the angle between the direction of ${\bf q}$ and the direction of $\theta =0$; the ellipsis again denotes analytical terms.
 We see that the non-analytical term is 
explicitly independent of the cutoff, confirming the consistency of our analysis.
An alternative evaluation of  $\delta \chi_{LW}(q)$ is presented in  the Appendix ~\ref{app:compl}.

For a circular Fermi surface, $k_0 = k_F$, $v_F$ is a constant, and 
Eq. (\ref{chi3}) reduces to the previously known result~\cite{Chubukov05}.
However, the previously published computations are arranged in a way which
apparently does not invoke the curvature at all. In Appendix C we show that 
the previous method does in fact involve the curvature, and leads to results
equivalent to those presented here.

\subsection{$2p_F$ processes}

To evaluate the contribution of $2p_F$ processes we could proceed from Eq (\ref{chibasic})
but expanding $\Lambda$ in 
${\tilde q} = q-Q$, where, we remind, 
 $Q =2k_{F}{\hat q}$.   As before the products of $G$ produce
an expression with all poles in the same half plane. Exploiting the non-analyticity of $\Pi({\bf q}+{\bf {\tilde q}},\Omega)$
we obtain an expression which has a nonanalytic part which  evaluates to the same expression as Eq (\ref{chi3}). 
Instead of presenting the details of this calculation, we present an alternative approach due to 
Belitz, Kirkpatrick, and Vojta~\cite{bkv},
 in which one partitions the diagram into two "triads", 
using the explicit form of $\Lambda_k$
\begin{equation}
\delta \chi (q) =-4\sum_{ab}\int(dl)(dk_1)(dk_2)\Gamma^{aabb}(\theta)^2\left[G^a(k_1+q)G^a(k_1)G^a(k_1+l)\right]
\left[G^b(k_2+q)G^b(k_2)G^b(k_2+l)\right]
\label{chitriad}
\end{equation}
where $\theta$ is the angle between ${\bf k}_1$ and ${\bf k}_2$. Choosing a particular point on the 
Fermi surface and evaluating the integral over $\varepsilon_{k_1}$ and
the associated frequency yields ($q_\Vert$ is the component of ${\bf q}$ parallel to the direction chosen for $k_1$)
\begin{eqnarray}
\delta \chi& &=-4\int\frac{k_0(\theta_1)d\theta_{1}}{2\pi (v^a)^2}\frac{\Omega}{({\bf v^a}\cdot {\bf q})}
\left(\frac{1}{(\frac{i\Omega}{v^a}-l_\Vert)}-\frac{1}{\frac{i\Omega}{v^a}
-{\hat {\mathbf v}}\cdot {\mathbf q }-l_\Vert)}\right)
\nonumber \\
&&\Gamma^{aabb}(\theta_1)^2\left[G^b(k_2+q)G^b(k_2)G^b(k_2+l)\right]
\label{chitriad2}
\end{eqnarray}
As in the previous calculation, singular contributions can only come from regions where  $k_2$ 
is directed oppositely to $k_1$. Choosing as origin of $k_2$ the point diametrically
opposite $k_1$, introducing parallel and perpendicular components as before and integrating over $k_{2\Vert}$, 
the associated frequency, and $l_\Vert$ we get 
\begin{eqnarray}
\delta \chi&=&\sum_{ab}4i \int\frac{k^a_0(\theta_1)d\theta_{1}}{2\pi (v^a)^2(v^b)^2}\frac{\Gamma^2 (\pi)}{(2\pi)^2}  T\sum_\Omega 
\int\frac{dk_\bot dl_\bot}{(2\pi)^2}\left(\frac{\Omega^2 \text{sgn} \Omega}{\left|({\bf v^a}\cdot {\bf q})({\bf v}^b \cdot {\bf q})\right|}\right) 
\nonumber \\
&&\left(\frac{1}{\frac{2i\Omega}{v_{avg}}-{\hat {\mathbf v}}\cdot {\bf q}-\frac{l_\bot k_\bot}{k^b_0}}
+\frac{1}{\frac{2i\Omega}{v_{avg}}+{\hat {\mathbf v}}\cdot {\bf q}
-\frac{l_\bot k_\bot}{k^b_0}}-\frac{2}{\frac{2i\Omega}{v_{avg}}-\frac{l_\bot k_\bot}{k^b_0}}\right)
\label{chitriad3_1}
\end{eqnarray}
Eq (\ref{chitriad3_1}) is seen to be of precisely the same form as Eq (\ref{chitriad3}) and gives the same result;
the only difference is the dependence on orbital index. 
Integrating  over frequency and combining the results 
from small $q$ and $2k_F$ contributions 
gives
an answer whose orbital dependent part depends on the velocities via the 
combination
\begin{equation}
\frac{v^a(v^b)^3 +(v^a)^3v^b+2(v^a)^2(v^b)^2}{(v^a+v^b)^3}\left( {\hat {\mathbf v}}\cdot {\mathbf q}\right)
=v_{avg} {\hat {\mathbf v}}\cdot {\bf q}
\end{equation}

\subsection{Final Result}

Collecting  the small $q$ and $2k_F$ contributions from all diagrams in Fig.~\ref{diagramsforchi} we find 
for the spin susceptibility
\begin{equation}
\delta\chi_s(q) =\sum_{ab} \frac{v_{avg}|q|}{6\pi^{3}}~
\int_{0}^{2\pi} \frac{k^a_0(\theta_k) k^b_0 (\theta_k)d \theta_k}{2\pi (v^a)^2(v^b)^2} (\Gamma^{ab,s} (\pi))^2
\left|{\hat{\mathbf v}}\cdot {\hat {\mathbf q}}\right|
 \label{eq32}
\end{equation}
At $q=0$ and $T>0$, we have
\begin{equation}
\delta\chi_s(T) = \sum_{ab} \frac{T}{\pi^{3}}
\int_{0}^{2\pi} \frac{d \theta_kk^a_0(\theta_k) k^b_0 (\theta_k)}{2\pi (v^a)^2(v^b)^2} (\Gamma^{ab,s} (\pi))^2). \label{eq33}%
\end{equation}
For an isotropic Fermi surface and one band, this again
reduces to the result in \cite{Chubukov05}.
To make contact with previous work, which considered a 
simplified interaction with indentical spin and charge components,
 $\Gamma^{ab,s} (\pi)$ has to be replaced by $\Gamma^{ab} (\pi)/2$.
Higher order powers of $\Gamma$ do contribute
 to $|q|$ and $T$ terms in $\chi_s$, in distinction to $C(T)/T$, and at strong coupling, the 
 non-analytic terms in the spin susceptibility are not expressed entirely 
 via $\Gamma^2_s (\pi)$~\cite{ch_masl_latest}. the terms of order $\Gamma^3$ have
 a less singular dependence on the curvature.
 At weak and moderate coupling, though, Eqs.
(\ref{eq32}) and (\ref{eq33}) should be sufficient. Finally, for 
the charge susceptibility $\chi_c (q,T)$  non-analytic contributions from individual diagrams are all cancelled out:
the full $\chi_c (q,T)$ is an analytic function of both arguments. 

\section{Fermi surfaces with Inflection Points}

\label{results}

We see from Eq. (\ref{eq32}) and (\ref{eq33}) that as long as $k_0$ is
finite all along the Fermi surface, the anisotropy of the Fermi surface
affects the prefactors for $|q|$ and $T$ terms, but do not change the
functional forms of the non-analytic terms in the specific heat and 
spin susceptibility.
New physics,
however, emerges when the Fermi surface develops inflection points at which
$k_0(\theta_k)$ diverges. 
Inflection points are a generic feature of realistic Fermi surfaces
of two dimensional materials. In this section we show how inflection points
emerge and then indicate the modifications they make to the results 
 presented above. 

\subsection{Inflection points in commonly occurring models}

We first note that many quasi-one dimensional organic materials
have a band dispersion described by 
\begin{equation}
H_{organic}=-2ta\left(|k_x|-k_{F}\right)-2t'cos(k_yb)
\label{Horganic}
\end{equation}
with $|t'|<<t$, lattice constants $a,b$ not too different, 
and a third dimensional coupling weaker than $t'$ by an order of magnitude. In this case 
$k_0=2t'cos(k_yb) +{\cal O}(t'^2/t)$
obviously vanishes at $k_yb\approx \pm \pi/2$. Thus inflection points are
generic to quasi one dimensional materials.

We now consider the fully two dimensional $t-t'$ model, with quasiparticle dispersion
\begin{equation}
\varepsilon_{k} = -2t (\cos k_{x} + \cos k_{y}) + 4 t^{\prime}\cos k_{x} \cos
k_{y} - \mu\label{d1}%
\end{equation}
We assume that $t$ and $t^{\prime}$ are positive and $\mu$ is negative;
$t$ should be larger than $2 t^{\prime}$ for stability.

Consider now the Fermi surface crossing along the $(0,0)\rightarrow (\pi,0)$ direction (if it exists).
The crossing occurs at 
\begin{equation}
-(2t-4t')cos(k_x)=\mu+2t
\end{equation}
the velocity is along $x$ and the curvature may be read off by expanding to second
order in $k_y$, giving
\begin{equation}
k_0({\hat x})=\frac{1}{(2t-4t'cos(k_x))}
\label{k0x}
\end{equation}
which is manifestly positive.

On the other hand, at the Fermi surface crossing along the diagonal $k_x=k_y$
we find
\begin{equation}
k_0({\hat x}+{\hat y})=\frac{1}{2tcos(k_x)-4t'}
\label{k0diag}
\end{equation}
which is positive at small $k_x$ but changes sign as the Fermi surface approaches the point $(\pi/2,\pi/2)$.
We therefore conclude that for chemical potentials in the appropriate range inflection points must exist
because the curvature has opposite  sign at two points on the Fermi surface.

\subsection{spin susceptibility and specific heat}

We now analyze how the inflection points affect the nonanalytic terms in the spin
susceptibility and specific heat.
For simplicity, we restrict to one-band syatems.
 Quite generally, near each of the inflection
 points $k_0(\theta)$ behaves
as
\begin{equation}
k_0(\theta) \propto(\theta- \theta_{0})^{-1}%
\end{equation}
At $\theta= \theta_{0}$, the curvature diverges, i.e., there is no quadratic
term in the expansion of the quasiparticle energy in deviations from the Fermi
surface. In the generic case, the dispersion is  then
\begin{equation}
\varepsilon_{k} = v_{F} (\theta_{0}) k_{\Vert} + A k^{3}_{\bot} \label{eq34}%
\end{equation}
where, as before, the directions $k_\Vert$ and $k_\bot$ 
are along and transverse to the
direction of the Fermi velocity at $\theta= \theta_{0}$. For a special
situation when $\theta_{0}$ coincides with a reflection symmetry axis for $\varepsilon
_{k}$, i.e., when $\mathbf{v}_{F} (\theta_{0})$ is directed along the
Brillouin zone diagonal in $t-t^{\prime}$ dispersion, the expansion of
$\varepsilon_{k}$ in the direction transverse to the zone diagonal holds in even
powers of $k_{\bot}$, i.e., at $\theta= \theta_{0}$,
\begin{equation}
\varepsilon_{k} = v_{F} (\theta_{0}) k_{\Vert} + B k^{4}_{\bot} \label{eq35}%
\end{equation}

In both cases, a formal integration over $\theta$ in Eqs. (\ref{eq32}) and
(\ref{eq33}) yields divergences. The divergences are indeed artificial and are
cut by either $A$ or $B$ terms in the dispersion. The effect on $\delta
\chi(q,T)$ and $\delta C(T)/T$ 
 can be easily estimated if we note that the angle integrals diverge
as 
$\int d \theta k^2_0 (\theta) \sim \int d \theta/(\theta-\theta_{0})^{2}$. In a generic case, described by
(\ref{eq34}), $1/|\theta- \theta_{0}|$ has to be replaced by $1/[|\theta-
\theta_{0}| + A k_{\bot,typ}]$.
The angle integral then yields $1/ k_{\bot,typ}$. It follows from Eq.
(\ref{eq34}) that $k_{\bot, typ} \sim(k_{\Vert,typ})^{1/3}$. 
It also follows from our consideration above that typical values of $k_{\Vert}$
 are of order $|q|$.
Combining the pieces, we find that the integral diverges as $|q|^{1/3}$, or
\begin{equation}
\delta\chi(q) \propto \Gamma^{2} (\pi)  |q|^{2/3} \label{eq36}%
\end{equation}     
Similarly, at finite $T$ we obtain
\begin{equation}
\delta\chi(T) \propto \Gamma^2 (\pi)  |T|^{2/3} \label{eq37}%
\end{equation}
And the specific heat
\begin{equation}
\delta C(T)/T \propto \Gamma^2 (\pi)  |T|^{2/3} \label{eq37_a}%
\end{equation}

For special inflection points along symmetry direction, the analogous
consideration shows the angle integral yields $1/(k_{\bot,typ})^{2}$. At the
same time, it follows from Eq. (\ref{eq35}) that $k_{\bot,typ} \sim
(k_{\Vert,typ})^{1/4} \sim|q|^{1/4}$. Then the angle integral diverges as
$|q|^{1/2}$, and
\begin{equation}
\delta\chi(q) \propto \Gamma^2 (\pi) |q|^{1/2} \label{eq36'}%
\end{equation}
while
\begin{equation}
\delta\chi(T) \propto \Gamma^2 (\pi) |T|^{1/2} \label{eq37'}%
\end{equation}
and 
\begin{equation}
\delta C(T)/T \propto \Gamma^2 (\pi)  |T|^{1/2} \label{eq37'_a}%
\end{equation}
The results, Eqs. (\ref{eq36} -(\ref{eq37'_a})), differ from the results by
Fratini and Guinea~\cite{Fratini02}. They obtained $\chi(T) \propto T \log T$
for a generic inflection point, and $\chi(T) \propto T^{3/4} \log T$ for a
special inflection point. In our calculations, a similar $T \log^{2} T$
behavior for a generic case would hold if the non-analytic terms were
coming from 
vertices $\Gamma (\theta)$ with 
arbitrary $\theta$ rather than $\theta = \pi$. 
Then, e.g., 
the coefficient for the $|q|$ term in the
 spin susceptibility would be given by a double
integral $\int d \theta_{1} |k_0(\theta_{1})|~\int d \theta_{2} |k_0 
(\theta_{2})|$. Each integral diverges logarithmically, 
and the correction would then 
scale as $T \log^{2} T$. The emergence of the anomalous power of temperature
or momenta for a generic inflection point in our calculations is the direct
consequence of the one-dimensionality of the relevant interaction. We see
therefore that the anisotropy of the Fermi surface is an ideal tool to probe
the fundamental 1D nature of the non-analyticities
 in a  Fermi liquid.

 \section{Application to $Sr_2RuO_4$}

 A crucial and so far unresolved question related to the results reported here
and in previous papers is the observability of the effects. Evidence
for the $T^3lnT$ nonanalyticities expected in three dimensional materials
have been observed in the specific heat of $^3He$ \cite{Abel66} 
(indeed this observation played a crucial role in stimulating the theoretical literature)
and similar effects have been noted in the specific heat of the heavy fermion material
$UPt_3$ \cite{Stewart84}.  
More recently, a linear in $T$ behavior has been observed in the specific heat 
of  fluid monolayers He$^{3}$
adsorbed on graphite~\cite{casey}, but to our knowledge no evidence for nonanalytic terms
in the susceptibility has been reported.

We consider here $Sr_2RuO_4$,
a highly anisotropic layered compound
for which detailed information about the shape of the Fermi surface,
the quasiparticle mass enhancements, the susceptibility
and optical conductivity is available \cite{Bergemann03,Ingle05,Lee06}. 
These data imply \cite{Bergemann03} that the material is "strongly correlated",
in the sense that Fermi velocities and susceptibilities are substantially renormalized
from the predictions of band theory. The data
also suggest that the dynamical self energy is at most weakly momentum-dependent,
because the shapes of the Fermi surface deviate only slightly from those found in band structure
calculations, implying that the self energy has a much stronger frequency dependence
than momentum dependence.  We extract
Fermi surface shape and Fermi velocities from quantum oscillation data \cite{Bergemann03}  
and cross-check with photoemission
data where available. We use available susceptibility \cite{Bergemann03}  and optical data
\cite{Lee06} to estimate the interaction functions.

It is also useful to consider the leading low-T behavior of the specific heat coefficient, which 
in a Fermi liquid  is given in terms of fundamental constants and
a sum over bands of the average of the inverse of the Fermi velocity
as
\begin{equation}
\frac{C}{T}=\frac{2\pi^2}{3} \sum_\lambda I_\lambda
\label{Cfl}
\end{equation}
with
\begin{equation}
I_\lambda=\oint \frac{d \theta}{4\pi^2} \frac{k_{F,\lambda}(\theta)\sqrt{1+\left(\frac{dk_F(\theta)}{k_F(\theta)d\theta}\right)^2}}{\left|v_{F,\lambda}(\theta)\right|}
\label{Idef}
\end{equation}

For Galilean-invariant fermions, it is customary to use the relation (we use units where $\hbar=0$)
$v_F=k_F/m$ to define a band mass 
\begin{equation}
m_\lambda=2\pi I_\lambda
\label{mlambdadef}
\end{equation}
so $C/T=(\pi/3)\sum_\lambda m_\lambda$. To obtain the specific heat in conventional units 
($mJ/mol/K^2$) 
one must multiply by $k_B^2$ and by the Avogardo
number.

We begin with the results for the Fermi surface shape and Fermi velocities. 
In $Sr_2RuO_4$, the relevant electrons are the three $t_{2g}$ symmetry 
$Ru$ d-orbitals and there are accordingly three
bands at the Fermi surface, conventionally labeled as
$\alpha,\beta,\gamma$. The Fermi surface {\em shape}, shown in Fig \ref{fsfig}
is well described by the two dimensional  tight binding model
\begin{eqnarray}
\varepsilon_{\gamma}(k_x,k_y)&=&-2t_{1\gamma}\left(cos(k_x)+cos(k_y)\right)-4t_{2\gamma}\left(cos(k_x)cos(k_y)\right) - \varepsilon_{0\gamma} \label{egamma} \\
\varepsilon_{\alpha,\beta}&=&-(t_{1\alpha} + t_{2\alpha})\left(cos(k_x) + cos(ky)\right) - \varepsilon_{0xz} 
\nonumber \\
&&\pm \sqrt{\left((t_{1\alpha} - t_{2\alpha})(cos(k_x)- cos(k_y)\right)^2 + 16t_{3\alpha}^2sin(k_x)^2sin(k_y)^2}
\label{eab}
\end{eqnarray}
(couplings in the third dimension are an order of magnitude smaller).

A detailed quantum oscillation study has been performed by Bergemann and collaborators \cite{Bergemann03}.
These authors present in Table 4 of their work 
a tight-binding parametrization which reproduces the {\em shape} of the Fermi surface. They
also present results for the mass enhancements in each Fermi surface sheet, which may be converted
into experimental estimates for $I_\lambda$.
The shape, of course, does not depend on the magnitudes of the tight binding parameters. We accordingly
rescale these in order to obtain velocities (more precisely, integrals $I_\lambda$) corresponding to the 
data reported by Bergemann et al\cite{Bergemann03}.

\begin{table}[htdp]
\caption{Tight binding band parameters (in [eV]) which reproduce the shape and, approximately,
the Fermi velocities of the three bands at
the Fermi surface of $Sr_2RuO_4$. Parameters are taken from
Table $4$ of Ref \cite{Bergemann03} and then renormalized to produce sheet-dependent
quasiparticle mass enhancements
approximately consistent with experiment. Last column: mass parameter
computed using Eqs (\ref{Idef}), (\ref{eab}).}

\begin{center}
\begin{tabular}{|clclclclclcl}
\hline
band&$\varepsilon_0$&$t_1$&$t_2$&$t_3$&$\frac{m_\lambda}{m_e}$&\\
\hline
$\alpha$&0.13&0.13&0.013&0.02&2.5&\\
$\beta$&0.16&0.15&0.013&0.02&5.8&\\
$\gamma$&0.012&0.079&0.032&0&16&\\ 
\hline
\end{tabular}
\end{center}
\label{default}
\end{table}%

A few remarks about the velocities and masses are in order. First, the calculated $\gamma$ band properties 
depend very sensitively
on how close the $\gamma$ ($xy$)-derived band approaches the van Hove points $(\pi,0)$, $(0,\pi)$. 
Published band calculations  \cite{Oguchi95,Mazin97,Liebsch00} show wide variations in the position of the
the singularity relative to the Fermi level. Second,  the $'mass'$ derived from the specific heat 
involves both the velocity and the geometrical
properties of the Fermi surface. The mass for the $\alpha$ band is small 
because of its small size, even though
its velocity is relatively small.  Third, and most important, the curvature of the $\alpha,\beta$
bands depends very sensitively on the parameters $t_{2\alpha},t_{3\alpha}$; the velocities also
depend somewhat on these parameters. 
A recent angle-resolved photoemission
experiment \cite{Ingle05} reports that the $\alpha$ band Fermi velocity at 
the zone face crossing point is
$v_\alpha=1.02eV-\AA$; the parameterization used here gives 
an essentially identical value.

We now turn to the Landau interaction function.  A complete experimental determination
is not available, but considerable partial information exists. Bergemann and co-workers \cite{Bergemann03}
have determined, for each band, the spin polarization induced by a uniform external magnetic
field, so the $"L=0"$ spin channel Landau parameters may be estimated.
Optical conductivity data \cite{Lee06} provide some information on the $"L=1"$ spin-symmetric
channel current response.  General arguments suggest that the charge compressibility
is only weakly renormalized in correlated oxide materials, allowing a rough estimate
of the $"L=0"$  charge channel interaction. We will use this information
to estimate the scattering amplitudes and hence the nonanalytic terms
in the susceptibilities. These estimations are certainly subject to large uncertainties,
but we hope they will give a reasonable idea of the magnitude of the effects. 

In the three-band material of present interest the interaction function
is a symmetric $3\times3$ matrix with components  $\Gamma^{a,b}$
labelled by orbital or band indices, which should then be decomposed into charge
(symmetric) and spin (antisymmmetric) components and into the angular harmonics
appropriate to the tetragonal symmmetry of the material.  We begin with the
isotropic $"L=0"$ spin channel. We assume (consistent with the usual practice
in transition metal oxides) that the deviations from $O(3)$ symmetry,
while crucial for electronic properties such as the band structure and conductivity,
are not crucial for the local interactions, which arise from the physics of the spatially
well localized $d$ electrons.  This implies that the interactions are invariant
under permutations of orbitals, so that it is reasonable to assume that  the two-particle
irreducible spin channel interaction  takes the simple Slater-Kanamori
form with two parameters, which we write as
$\Gamma^{a,a}\equiv U_{eff}$ and $\Gamma^{a\neq b}\equiv J_{eff}$. 
Thus the physical static susceptibilities are given by
\begin{equation}
{\mathbf \chi}=\left({\mathbf \chi}^{-1}_0+{\mathbf \Gamma}(U_{eff},J_{eff})\right)^{-1}
\label{chidef}
\end{equation}
and  fix the parameters $U_eff$ and $J_{eff}$ by comparing measured susceptibilities
to the values predicted by the renormalized tight binding parameters.

Ref \cite{Bergemann03} presents  (as mass enhancements)
data for the spin susceptibility of each band (obtained from the spin splitting of the Fermi surfaces);
finding $\chi^{\alpha,\alpha}/\chi_0^{\alpha,\alpha}\approx 1.2$, 
$\chi^{\beta,\beta}/\chi_0^{\beta,\beta}\approx 1.3$ and  
$\chi^{\gamma,\gamma}/\chi_0^{\gamma,\gamma}\approx 1.6$. We 
estimate $U_{eff}\approx 0.033$ and $J_{eff}=-0.008$, where $\chi_0$ is the 
susceptibility implied by the quantum oscillations Fermi surface and
mass. This implies that the dimensionless Landau interaction parameters
(in the limit $\omega/k\rightarrow 0$ limit 
$A^{ab}\equiv \Gamma^{ab}\sqrt{ \chi^{a}\chi^b}$
\begin{equation}
{\mathbf A}^{spin}=\left(\begin{array}{ccc}-0.053& 0.040 & 0.10 \\0.04 &- 0.13& 0.17 \\0.10 & 0.17 & -0.40\end{array}\right)
\label{gamanti}
\end{equation}

The uncertainties in the off diagonal components are large, perhaps $50\%$,
but because the interactions enter squared, the contribution of the off diagonal
components is not very significant.
The  dominant term is the  $\gamma-\gamma$ band interaction, as expected because
it has the largest mass and the largest susceptibility enhancement, but that all of the other
contributions taken together make a non-negligible contribution to the interaction.
Finally, we note that the spin channel renormalizations are not large, so use of the second order result
is not unreasonable. 

We now turn to the charge channel, beginning with the compressibility.
There is no experimental information available. However, it is generally believed that for systems, 
such as transition metal oxides, with strong local interactions the total charge susceptibility is not strongly
renormalized, so that the Landau parameter acts to undo the effects of the mass enhancement.
Further, if the $J$ (orbital non-diagonal) component of the interaction is not too small
relative to the $U$ (orbital diagonal component) then a residual interaction acts to 
shift the levels such that the ratio of occupancies of each of the three $t_{2g}$ orbitals remains
constant under chemical potential shifts.
Taking as unrenormalized value the susceptibilities folllowing from the tight binding parameters given
in Ref \cite{Bergemann03} we then obtain
\begin{equation}
{\mathbf A}^{S,0}=-\left(\begin{array}{ccc}0.40 & 0.034 & 0.026 \\0.034 & 0.8 & 0.026 \\0.026 & 0.026 & 0.80\end{array}\right)
\label{Fc}
\end{equation}
with again considerable uncertainty in the off diagonal components.

Finally, we turn to the current renormalization.
The optical conductivity is commonly presented in the extended Drude form
\begin{equation}
\sigma(\Omega)=\frac{\frac{e^2}{\hbar c}D_{band}}{-i\Omega\frac{m^*(\Omega)}{m}+\Gamma(\Omega)}
\label{drudedef}
\end{equation}
where $c$ is the mean interplane spacing and 
 $m^*/m$ has 
the meaning of an optical mass
enhancement defined with respect to a reference value determined by $D_{band}$.
In a Fermi liquid at low temperatures, $\Gamma(\Omega \rightarrow 0)$ is very small
and  (assuming tetragonal symmetry)
\begin{equation}
D_{band}\frac{m}{m^*(0)}\equiv D=\sum_\lambda
\oint \frac{d \theta}{4\pi^2} k_{F,\lambda}(\theta)
\sqrt{1+\left(\frac{dk_F(\theta)}{k_F(\theta)d\theta}\right)^2}
\left|v_{F,\lambda}(\theta)\right|\left(1+\frac{F^{1S}_\lambda}{2}\right)
\label{ddef}
\end{equation}
Note that the numerical value of the mass enhancement $m^*/m^0$ depends on the choice of
reference value $D_{band}$ but that $D$ is a physically meaningful quantity determined
directly from the data. 

The room temperature conductivity of $Sr_2RuO_4$ has been measured \cite{Lee06}.
These authors chose the value $e^2D_{band}/\hbar c$ 
(which they denote as $\omega_p^2/4\pi$) to correspond to 
$\omega_p^2\approx 8\times 10^{8} cm^{-2}$ and 
find $m^*/m(\Omega \rightarrow 0)$ (which they denote
as $\lambda$) to be $\approx 3.5$.
This implies that $D\approx 0.13eV$, somewhat smaller
than the value $0.18$ obtained from
Eq \ref{ddef}, implying that the average over all bands is
$F^{1S} \approx -0.55$. The temperature dependence of $D$ in $Sr_2RuO_4$
has not been measured, but it seems 
reasonable that $D$ should decrease as $T$
decreases, implying a  further increase in the magnitude of $F^{1S}$. Determining the 
temperature dependence of the optical mass is therefore
an important issue.

Ref \cite{Bergemann03} presents, as masses, data for the cyclotron resonance frequencies
for the different Fermi surface sheets. 
These masses should be essentially equivalent
to the $D$ values quoted above. 
Bergemann et. al.  emphasize that the frequencies
are subject to large errors, and that the results should be regarded as tentative.
The quoted 
cyclotron  masses correspond to $D$ values about a factor of two larger
than those implied by the measured fermi velocities, and about a factor
of $3$ larger than the values inferred from the optical data. In view
of the stated large uncertainties in the measurement and the qualitative 
incosistency with the optical data, we
disregard the cyclotron resonance measurements here. 

Now, the crucial object for the specific heat is the backscattering amplitude.
A negative  $F^{1S}$ implies a positive backscattering amplitude, so 
as a rough approximation to the effects of the current channel Landau renormalization
we add the interaction corresponding to 
$F^{1S}=-0.6$ to the diagonal components of Eq \ref{Fc}.

The crucial points emerging from this analysis are that the 
reducible interactions  in the charge channel are of order unity, 
whereas those in the spin channel are somewhat smaller,
implying a larger nonanalyticity in the specific heat than in the susceptibility.
Substituting the  interaction amplitudes into
Eqs. \ref{apr4_1_1}, \ref{eq33} and performing the fermi surface averages then yields the following estimates
\begin{eqnarray}
\gamma(T)&=&36 ~mJ/mol-K^2\left(1-0.0015T[K]\right) \\
\chi(T)[Si/Volume]&=&1.5\times 10^{-4} \left(1-.00001 T[K]\right)
\end{eqnarray}
The  small magnitude of the corrections (especially to the spin 
susceptibiltiy) follows from the small prefactors in Eq
(\ref{eq33}) and the not too large Landau renormalizations. 
The size of the effect is increased by the relatively small curvatures of the $\alpha$
and, especially, $\beta$ bands, and we note that substantial increases in the coefficients
occur if the mixing coefficients $t_{2\alpha},t_{3\alpha}$ in Eq \ref{eab} are reduced.
We expect the results to
be valid above a (still not well determined) scale probably $\sim 1-2K$ 
at which the Fermi surface
warping becomes important enough to make the material three dimensional,
and below the scale at which Fermi liquid theory breaks down,
and we see that temperatures of order $10K$ lead to $20\%$ deviations
in the value of the specific heat coefficient and to $1\%$ changes in $\chi$.

Replacing $Sr$ by $Ca$ leads to a dramatic (factor $\sim 100$ in $Sr_{0.5}Ca_{1.5}RuO_4$) 
enhancement of the susceptibility. It seems likely that this increase is
not due to a decrease in the fermi velocities, but must be interpeted
as a dramatic increase in the spin Landau parameter,
suggesting perhaps that nonanalytic $T$-dependence of $\chi$
might be more easily observed in $Ca$ doped materials, 
although in this case disorder effects
would need to be considered.

 \section{Conclusions}

In this paper, we studied non-analytic terms in  the spin susceptibility
 and specific heat in 2D systems with anisotropic, non-circular 
 Fermi surfaces. For systems with circular Fermi surfaces, the 
non-analytic terms in $\chi_{s} (q,T)$ and $C(T)/T$ 
are linear in $max (q,T)$. We
argued that the anisotropy of the Fermi surface serves as a testing ground to
verify the theoretical prediction that the non-analytic terms 
 originate from a single 1D scattering amplitude which combines two 1D
interaction processes for particles at the Fermi surface in which the
transferred momenta are either $0$ or $2k_{F}$, and, simultaneously, the total
moment is zero. We obtained explicit expressions for the non-analytic momentum
and temperature dependences of the spin susceptibility and the specific heat 
 in systems with non-circular  Fermi surfaces and demonstrated that for the
 Fermi surfaces with
inflection points, the the non-analytic temperature and momentum dependences
 are $\chi_{s} \propto max (q^{2/3}, T^{2/3})$, $C(T)/T \propto T^{2/3}$
 in a generic case, and as
$\chi_{s} \propto max (q^{1/2}, T^{1/2})$, $C(T)/T \propto T^{1/2}$
 for the special cases when the inflection
points are located along symmetry axis for the quasiparticle dispersion.
We estimated the order of magnitude of the effects in the quasi two dimensional
material $Sr_2RuO_4$.

It is our pleasure to thank C. Bergemann, D.L. Maslov, A. Mackenzie and N. Ingles  for useful
conversations. The research  is supported by NSF
Grant No. DMR 0240238 (AVC) and DMR 0431350 (AJM),
and AJM thanks the DPMC at the University of Geneva for hospitality
while this work was completed.

\appendix
\section{The details of the evaluation of Eq.(\ref{chi3})}
\label{app:details}
For simplicity, we neglect band index, i.e., set $v^a_k = v^b_k = v_k$, and $k^a_0 (k) = k^b_0 (k) = k_0 (k)$.
Using (\ref{12_1}), Eq. (\ref{chismallq2}) is re-expressed as
\begin{eqnarray}
&&\delta \chi_{LW}(q) =-\frac{\Gamma^2 (\pi)}{64\pi^5}
\int \frac{d\theta_k k_0 (k)}{2\pi v^3_k}\int_{-\Lambda}^\Lambda \int_{-\Lambda}^\Lambda \frac{dx dy}{
\epsilon^2}~\times
\nonumber \\
&& \left[(xy-\epsilon)^2 \log{(xy-\epsilon)^2} + (xy+\epsilon)^2 \log{(xy+\epsilon)^2} - 2 x^2 y^2 \log{x^2y^2}\right]
\label{12_2}
\end{eqnarray}
where $\epsilon = ({\bf v}_k \cdot {\bf q}) k_0 (k)/v_k$.  
Rescaling $x = \sqrt{|\epsilon|} {\bar x}, ~y = \sqrt{|\epsilon|} {\bar y}$, substituting into (\ref{12_2}) and dropping irrelevant 
terms confined to high energies, we obtain
\begin{equation}
\delta \chi_{LW}(q) = -\frac{\Gamma^2 (\pi)}{64\pi^5}
\int \frac{d\theta_k k_0 (k) |\epsilon|}{2\pi v^3_k} Z
\label{12_3}
\end{equation}
where
\begin{equation}
Z = \int_{-\Lambda}^\Lambda \int_{-\Lambda}^\Lambda 
d{\bar x} d{\bar y}  \left[({\bar x} {\bar y}-\epsilon)^2 \log{({\bar x} {\bar y}-\epsilon)^2} + ({\bar x} {\bar y}+\epsilon)^2 \log{({\bar x} {\bar y}+\epsilon)^2} - 2 ({\bar x} {\bar y})^2 \log{({\bar x} {\bar y})^2}\right]
\label{12_21}
\end{equation} 
Introducing further ${\bar x} = \sqrt{2r} \cos \phi/2$ and ${\bar y} = \sqrt{2r} \sin \phi/2$,
 we rewrite $Z$ as 
\begin{eqnarray}
Z = 2 \int_{0}^\pi d \psi  \int_{0}^\Lambda 
&dr& \left[(r \sin{\phi} -1)^2 \log{(r \sin{\phi} -1)^2} + 
(r \sin{\phi} +1)^2 \log{(r \sin{\phi} -1)^2} \right. \nonumber \\
&&\left. - 2 r^2 \sin^2{\phi} \log{ r^2 \sin^2{\phi}}\right]
\label{12_31}
\end{eqnarray}
Subtracting the irrelevant large $r$ contribution $6 + \log{ r^2 \sin^2{\phi}}$
 from the integrand in (\ref{12_31}), we the universal part of $Z$ in the form
\begin{eqnarray}
Z = 2 \int_{0}^\pi d \psi  \int_{0}^\Lambda 
&dr& \left[ r^2  \sin^2{\phi} \log \left(1 - \frac{1}{r^2 \sin^2{\phi}}\right)^2 
 + 2 r  \sin{\phi}  \log\left(\frac{1 + \frac{1}{r \sin{\phi}}}{1 -\frac{1}{r \sin{\phi}}}\right)^2 \right. \nonumber \\
&&\left. -6 + \log \frac{ (r^2 \sin^2{\phi} -1)2}{r^4 \sin^4{\phi}}\right]
\label{12_4}
\end{eqnarray}
One can make sure that the integral over $r$ vanishes if we set the upper
 limit at $\Lambda = \infty$. as the integrant depends on $r$ only via 
$r \sin \phi$, the finite contribution to the integral comes from $\lambda \sin \phi = O(1)$, i.e. from a narrow range of $\phi$ either near zero or near $\pi$. The contributions from these two regions are equal. Resticting with the 
 contribution from small $\phi$, expanding $\sin \phi \approx \phi$ and
 introducing $z = r \phi$ and $t = \Lambda \phi$, we obtain from (\ref{12_4})
\begin{equation}
Z = 4 \int_{0}^\infty \frac{dt}{t}  \int_{0}^t 
dz \left[ z^2 \log \left(1 - \frac{1}{z^2}\right)^2 
 + 2 z\log\left(\frac{1 + \frac{1}{z}}{1 -\frac{1}{z}}\right)^2  -6 + \log \left(\frac{z^2 -1}{z^2}\right)^2\right]
\label{12_5}
\end{equation}
Changing the order of the integration, we obtain for the universal part of $Z$
\begin{eqnarray}
Z &=& -4 \int_{0}^\infty dz \log z \left[ z^2 \log \left(1 - \frac{1}{z^2}\right)^2 
 + 2 z\log\left(\frac{1 + \frac{1}{z}}{1 -\frac{1}{z}}\right)^2  -6 + \log \left(\frac{z^2 -1}{z^2}\right)^2\right] \nonumber \\
&&= \frac{4\pi^2}{3}
\label{12_6}
\end{eqnarray}
Substituting this into (\ref{12_3}) and using the definition of $\epsilon$, we
 reproduce (\ref{chi3}).
  
\section{An alternative evaluation of  $\delta\chi_{LW} (q)$}
\label{app:compl}
In this Appendix we present a complementary 
 evaluation of $\delta\chi_{LW} (q)$ using a somewhat different 
 computational procedure. We again restrict to one band.
The point of departure are Eqs. (\ref{chibasic}) and (\ref{lambdadef}), 
 which we re-write at $T=0$ as 
\begin{align}
\delta\chi_{LW} (q)  &  = -4 \int\int\int\int\frac{d^{2}k~d^{2}q~d\omega
d\Omega }{(2\pi)^{6}}~ \Gamma^{2} \Pi(l,\Omega ) G_{0}(\mathbf{k}%
,\omega ) \times\nonumber\\
&  G_{0}(\mathbf{k}+\mathbf{l},\omega +\Omega )~ G_{0}(\mathbf{k}%
+\mathbf{q}+\mathbf{l},\omega +\Omega )~G_{0}(\mathbf{k}%
+\mathbf{q},\omega )
\label{23_1}
\end{align}

 We first integrate
over internal momenta $\mathbf{k}$ and frequency $\omega$ in the fermionic
 propagator. Expanding the result 
in $q^{2}$, we obtain
\begin{equation}
\delta\chi_{LW} (q) \propto \Gamma^2 q^{2} \int_{0}^{\infty}d \Omega
\Omega \int d^{2} l \int d \theta_{1} \frac{k_0 (\theta_{1})}{v_{F}
(\theta_{1})}~ \frac{1}{ (i \Omega - v_{F} l \cos\theta_{1})^{5}} \Pi(l,
\Omega) \label{eq27}%
\end{equation}
where $\theta_{1}$ is the angle between $\mathbf{l}$ and $\mathbf{k}$. 
 Directing $l_{x}$ and $l_{y}$ along and transverse to 
$\mathbf{k}$ and  substituting the polarization operator we obtain 
\begin{align}
\delta\chi_{LW} (q)  &  \propto \Gamma^2 q^{2} \int_{0}^{\infty}d
\Omega \Omega^{2}_{m} \int d l_{x} \int d \theta_{1} \frac{k_0(\theta_{1}%
)}{v_{F} (\theta_{1})}~ \frac{1}{ (i \Omega - v_{F} (\theta_{1})
l_{x})^{5}}~\times\nonumber\\
&  \int d l_{y} \int d \theta~\frac{k_0 (\theta_{1} + \theta)}{ v_{F}
(\theta_{1} + \theta)}\frac{1}{i\Omega - v_{F} (\theta_{1} + \theta)
(l_{x} \cos\theta+ l_{y} \sin\theta)} \label{eq28}%
\end{align}
[$\theta$ is the angle between two internal momenta 
 $\mathbf{p}$ and $\mathbf{k}$]. 
We now integrate over $l_{y}$ and then over $\theta$. The full result for this 2D
integral depends on particular forms of $k_0(\theta)$ and $v_{F} (\theta)$.
However, we only need 
  from the integral over $d l_y d \theta$ the term which is non-analytic  
 in the lower half-plane of $l_{x}$
 (this will allow us to avoid a  
degenerate pole at $v_{F} l_{x} = i \Omega$). 
 One can easily verify that the non-analyticity comes from the 
 integration near $\theta= \pi$ which  yields, instead of the
 second line in (\ref{eq28})
\begin{equation}
i ~\frac{k_0(\theta_{1}-\pi)}{2 v^{2}_{F} (\theta_{1}-\pi)} \log\left[
i\Omega + v_{F} (\theta_{1} -\pi) l_{x}\right]
\end{equation}
 For the Fermi surfaces with inversion symmetry (which we will only consider)
$k_0 (\theta_{1}-\pi) = k_0 (\theta_{1})$ and $v_{F} (\theta_{1}-\pi) = v_{F}
(\theta_{1})$ (we recall that $v_{F} (\theta)$ is the modulus of the Fermi
velocity at a particular $\theta$). Substituting this result into (\ref{eq28})
and extending the integral over $l_{x}$ onto the lower half-plane, 
we obtain Eqn (\ref{chi3}).

We also verified that the same result could be obtained by evaluating the
singular part of $\Pi(l, \Omega)$ by explicitly expanding near $\mathbf{p} = -
\mathbf{k}$ and expanding the dispersion $\varepsilon_{p} = \varepsilon_{-k + l}$ to
second order in $l$. In this computation, one power of $k_0(\theta)$ comes from
expanding the dispersion, while the other comes from the Jacobean of the transformation
from $d^{2}k$ to  $d\varepsilon_{k} d\theta$.

\section{Reevaluation of  $\delta\chi_{LW} (q)$ for an isotropic Fermi surface}
\label{app:B}

In this Appendix, we 
reconsider
a previously published \cite{Chubukov05} evaluation of  $\delta\chi_{LW} (q)$.
Although this evaluation leads to results identical to those we 
presented in the body of the paper
 for a circular Fermi surface,
 it apparently does not invoke the curvature explicitly.
Here we deconstruct this analysis, showing how the curvature actually enters
even when Fermi surface is circular.

We begin from Eq (\ref{23_1}). The analysis presented in the main body of the paper
involves choosing a direction for ${\bf l}$, and then performing the integral over ${\bf k}$,
which picked out points with a definite relationship to ${\bf l}$ and involved
the curvature in a direct way, and finally
integrating over $l$. On the other hand, the "conventional" analysis involves first fixing  the
direction of ${\bf k}$, integrating over the magnitude of $k$ and  over $q$, and then 
averaging over the direction of ${\bf k}$. In this method one
 expands $\epsilon_{k}$, $\epsilon_{k+l}$, $\epsilon
_{k+q}$ and $\epsilon_{k+l +q}$ in (\ref{23_1}) 
to linear order in the deviations from the
Fermi surface as $\epsilon_{k} = v_{F} (k-k_{F})$, $\epsilon_{k+l} = 
\epsilon_{k} + v_{F} l_{x}$, etc.  Because the Green functions have been linearized
the curvature apparently does not enter, in contrast to the previous derivation,
where the dependence of the Green function lines on curvature was essential. 

Integrating over $k$ and  over the
corresponding Matsubara frequency, and expanding the result in powers of $q$,
we obtain at $T=0$, neglecting regular terms
\begin{equation}
\delta\chi_{LW} (q) \propto \Gamma^{2}  q^{2} 
\int_{0}^{\infty}d \Omega
\Omega \int dl_{x} \frac{1}{ (i \Omega - v_{F} l_{x})^{5}} ~\int d
l_{y} \Pi({\bf l}, \Omega) \label{eq1}%
\end{equation}
The key point is that the curvature dependence is hidden in the polarizibility $\Pi$,
but in the circular Fermi surface limit this dependence is hidden.
To make the curvature dependence manifest 
 we use the fact that only backscattering contributes 
and evaluate the polarization bubble 
$\Pi(l, \Omega) = \int d^{2} t d \omega^{\prime}G_{0} (t, \omega^{\prime}) 
G_{0} (l+t, \Omega + \omega^{\prime})$ by expanding near 
$\mathbf{t} = -\mathbf{k}$. 
Introducing $\mathbf{t} + \mathbf{k} = \mathbf{p}$ 
and assuming that $p$ is small, 
we  expand the dispersions 
$\epsilon_{t} = \epsilon_{-k_x + \mathbf{p}}$ and $\epsilon_{t+l} = \epsilon_{-k_{x} + \mathbf{p} + \mathbf{l}}$
to second order in $p$: 
\begin{equation}
\epsilon_{t} = \epsilon_{-k_{x} + \mathbf{p}} = - v_{F} \left(p_{x} + \frac
{p^{2}_{y}}{2k_0}\right); ~~ \epsilon_{t+l} = \epsilon_{-k_{x} + \mathbf{p} +
\mathbf{l}} = - v_{F} \left(p_{x} + l_{x} + \frac{(p_{y} + l_{y})^{2}}{2k_0}
\right)
\label{eq6}%
\end{equation}
Substituting this expansion into the bubble and integrating over $p_{y}$ we
obtain
\begin{equation}
\Pi(l, \Omega) = i \frac{\Omega k_0}{2\pi^{2} v^2_{F} l_{y}}~\log{\frac{A
l_{y} - (i\Omega + v_{F} l_{x})}{-A l_{y} - (i\Omega + v_{F} l_{x})}}
\label{eq7}%
\end{equation}
where $k_0 A/v_F \sim k_F$ is the upper limit of the integral over $p_y$.
Integrating next $\Pi (l, \Omega)$ over $l_{y}$, 
we find  the same branch cut singularity as in ``conventional'' approach
\begin{equation}
\int dl_{y} \Pi(l, \Omega) = \frac{k_0 \Omega}{\pi v^2_{F}}~\log[i\Omega+
v_{F} l_{x}] \label{eq8}%
\end{equation}
Substituting this result into (\ref{23_1}) 
and using the fact that  $d^2 k$ in (\ref{23_1})
 can be re- expressed 
 as $(k_0/v_F) d \epsilon_k d \theta$, we 
 reproduce Eq. (\ref{chi3}) for a circular Fermi surface, 
and also reproduce Eq. (4.18) in [\onlinecite{Chubukov05}a], 
but with $k_0 /v_F$ instead of $m$.

For completeness, we also show
 that $\delta \chi_{2k_F} (q)$ in systems with 
 a circular Fermi surface can also be obtained with and without the curvature. 
A ``conventional'' computation~[\onlinecite{Chubukov05}a]
 expresses $\delta \chi_{2k_F} (q)$ in terms of the curvature. An alternative
 computational scheme  involves the same 
``triad'' method that we used in the main text.
In this scheme, the original expansion near $2k_F$ momentum transfer 
 contains the curvature, but it disappears from the answer at the latest stage.
 Performing the same  integrations over $\epsilon_k$, the corresponding frequency and $l_y$ as in the main text, we find (keeping $\Gamma = \Gamma (\theta)$)
\begin{eqnarray}
&&\delta\chi_{2k_{F}} (q) \propto\sqrt{k_0} \int d \theta \Gamma^{2} (\theta)  
\int_{v_F |q|}^{\infty}d\Omega
\Omega^{2}\nonumber \\
&& \times \int\frac{dl_{x}}{(l_{x} - i
\Omega)^{2} (i\Omega - v_F(l_{x} \cos\theta+ \frac{k_0}{2} \sin^{2}%
\theta))^{3/2}} \label{eq22}%
\end{eqnarray}
For $\cos\theta<0$, the two double poles are in different half-planes of
$l_{x}$. Integrating over $l_{x}$, we then obtain
\begin{equation}
\delta\chi_{2k_{F}} (q) \propto\sqrt{k_0} \int_{\pi/2}^{\pi} 
d \theta \Gamma^{2} (\theta)~\int_{v_F |q|}^{\infty} \frac{d\Omega\Omega^{2}}
 {(\frac{v_F k_0}{2}
\sin^{2}\theta+ i\Omega (1-\cos\theta))^{5/2}} \label{eq23}%
\end{equation}
Since relevant $\Omega \sim v_{F} |q|$, the $\theta$
integral is confined to $\theta=\pi$. Expanding near $\pi$ we obtain
\begin{equation}
\int_{\pi/2}^{\pi} \frac{d \theta \Gamma^{2} (\theta)}{(\frac{v_F k_0}{2} \sin^{2}\theta+
i\Omega (1-\cos\theta))^{5/2}} \approx \Gamma^{2} (\pi) \int_{0}^{\infty}
\frac{dx}{(\frac{v_F k_0}{2} x^{2} - 2 i \Omega)^{5/2}} = -\frac{U^{2}(\pi)}{3 \Omega^{2} \sqrt{v_F k_0}} \label{eq24}%
\end{equation}
Substituting this into (\ref{eq23}), we find that $k_0$ is 
canceled out, and
\begin{equation}
\delta\chi^{2k_{F}} (q) \propto \Gamma^{2} (2k_{F}) \int_{v_{F} |q|}^{E_{F}} d
\Omega\rightarrow \Gamma^{2} (\pi) |q|
\end{equation}
 Restoring the prefactor, we reproduce the same result as in the main text, but with $m v_F$ instead of $k_0$.


\begin{thebibliography}{99}
\bibitem{comm_2} The evaluation of the prefactor requires extra care. 
The dynamic part of $\delta \Pi_Q$ is 
nonanalytic only when $q_{\Vert} <0$, and
 for these $q_{\Vert}$, $(\delta \Pi_Q)^2 \propto \Omega^2/(-q_{\Vert})$.
At $q_{\Vert} >0$,  dynamic part of $\Delta\Pi_Q$ scales as $\Omega^2/(q_{\Vert})^(3/2)$, but
 static part scales as $\sqrt{q_{\Vert}}$.   The cross-product again gives
$(\delta \Pi_Q)^2 \propto \Omega^2/(q_{\Vert})$, and the prefactor is the same as at negative $q_{\Vert}$.
\end{thebibliography}
\end{document}